\documentclass[aps,pra,reprint,showpacs,amsfonts,amsmath,floatfix,superscriptaddress,footinbib]{revtex4-2}

\usepackage[svgnames]{xcolor}
\usepackage{glossaries}

\usepackage[colorlinks=true,        
            allcolors = black,  
            citecolor=DarkBlue, 
           linkcolor=DarkBlue,
            urlcolor=DarkBlue
        ]{hyperref}  
\usepackage{amsmath, amsfonts, amssymb, amsthm, bbm}

\usepackage{dsfont}
\usepackage{bm} 
\usepackage{mathtools}
\usepackage{physics}
\usepackage{graphicx}

\usepackage{dcolumn}

\usepackage{tikz}
\usepackage{float}

\usepackage{array,multirow}

\usepackage{bbold}
\usepackage{comment} 
\usepackage{array}
\usepackage{tabularx}

\usepackage{orcidlink}

\newcolumntype{C}[1]{>{\centering\arraybackslash}m{#1}}

\newcommand{\ad}{\hat{a}^\dagger}
\renewcommand{\a}{\hat{a}}
\newcommand{\n}{\hat{n}}

\newcommand{\nmax}{N_{\rm{max}}}

\usepackage{enumitem}
\newlist{todolist}{itemize}{2}
\setlist[todolist]{label=$\square$}
\usepackage{pifont}

\makeatletter
\def\switch@array{}
\makeatother

\begin{document}

\title{Analog quantum simulation of bosonic and anyonic models with flux-driven transmons}

\author{Isak Lyngfelt \orcidlink{0009-0008-1106-3038}}%
\email{isak.lyngfelt@chalmers.se}
\author{Jorge Fernández-Pendás \orcidlink{0000-0003-3814-9364}}
\author{Göran Johansson \orcidlink{0000-0001-5193-5250}}
\author{Laura García-Álvarez \orcidlink{0000-0002-3367-8083}}
\affiliation{%
Department of Microtechnology and Nanoscience (MC2), Chalmers University of Technology, SE-412 96 G\"{o}teborg, Sweden
}

\begin{abstract}
When quantum particles interact, many-body phenomena that are hard to simulate classically emerge. Quantum analog simulation offers an alternative in which the target system's dynamics is directly realized in controllable quantum hardware. Here, we give a general protocol for simulating the Bose–Hubbard and anyon-Hubbard models using lattices of capacitively coupled flux-tunable transmons. By modulating the transmon frequencies in an alternating pattern, we resonantly drive multiple many-body transitions and can tune the on-site interaction and density-dependent hopping amplitudes for up to three bosons per site, with no additional restriction on the total particle number. By adding phases to the modulation, which renders the transition amplitudes complex-valued, we propose the first simulation protocol for the anyon-Hubbard model with transmons. Numerical simulations of the driven transmon arrays with experimentally realistic parameters reproduce the characteristic dynamics of the target models across a range of interaction strengths and statistical phases, including the interaction-dependent localization and the statistics-dependent asymmetry of the anyonic quantum walk.
\end{abstract}
\maketitle

\section{Introduction}

Quantum hardware has undergone rapid technological progress in recent years, with several platforms transitioning from noisy intermediate-scale devices to early fault-tolerant regimes~\cite{acharya2023suppressing,bluvstein2024logical,bluvstein2026a-fault-tolerant}. Despite this improvement, the realization of practical computational advantage remains an open challenge~\cite{huang2025the-vast,babbush2026grand}. Whereas much effort has focused on error correction and algorithmic development, quantum simulations offer an alternative use of this hardware originally designed for gate-based quantum computation~\cite{altman2021quantum,daley2022practical}. Digital simulation uses gates to reproduce the model dynamics, which allows more flexibility in the target models at the cost of errors arising from product approximations of the evolution. In contrast, analog simulation does not rely on explicit qubit encodings or gate operations. Instead, the intrinsic dynamics of the device are used to realize a target model.

The motivation for quantum simulation lies in the expectation that these platforms enable access, at sufficient scales, to regimes of many-body physics that are difficult to simulate classically~\cite{altman2021quantum,daley2022practical}.
Such experiments aim to capture qualitative features of many-body systems, even when precise quantitative predictions are difficult.
However, a central challenge to this practical utility is the simulators' verification and validation, as a direct comparison with classical calculations would not be possible. 

In this context, recent work~\cite{trivedi2024quantum} addresses the advantage of noisy analog quantum simulators by considering the stability of relevant observables against errors. A complementary verification approaches is cross-platform comparison~\cite{altman2021quantum}, where the same model would be implemented in different architectures. Moreover, certain observables may be easier to measure in some platforms than others, providing access to different properties of the system. 

Beyond the study of many-body physics, using quantum processors for simulation contributes to their characterization and serves as a tool to probe the underlying hardware. The response of the system to controlled interactions can be used to test whether effective models remain valid for increasing sizes, identify limitations, and refine calibration strategies.

Several architectures can be used as analog simulators, including ultracold atoms in optical lattices, trapped ions, quantum dots, nanophotonic systems, and superconducting circuits. While the present work focuses on superconducting circuits, optical lattices provide an important reference point because they realize the target models considered in this paper. In particular, optical lattices have enabled analog quantum simulations of the Bose-Hubbard (BH) model, which captures key features of interacting bosonic systems and allows the study of superfluid--Mott-insulator transitions and lattice dynamics with a small set of controllable parameters (hopping amplitude, on-site interaction, and chemical potential)~\cite{cazalilla2011one-dimensional,chanda2025recent}. These models also admit generalizations with fractional exchange statistics. For example, the anyon-Hubbard (AH) model has been studied in optical lattices~\cite{keilmann2011statistically, greschner2015anyon, kwan2024realization}.

The physics of interacting bosons has also been explored with superconducting circuits~\cite{ticea2025observation, wang2026observing, mansikkamaki2022beyond, zhang2025synthetic, castillo-moreno2026experimental}. Superconducting architectures, such as the transmon qubit, are among the leading candidates for quantum computation, with current systems operating at the scale of a hundred qubits~\cite{acharya2023suppressing,kim2023evidence}. However, the transmon has unevenly spaced energy levels, with an anharmonicity that is typically much larger than the capacitive coupling between transmons, such that each mode is often reduced to the two lowest energy levels, i.e., a qubit. Therefore, the hard-core BH model and spin systems are popular target models for quantum analog simulation with superconducting qubits~\cite{rosen2025flat-band, rosen2024a-synthetic, guo2021stark, xu2018emulating, praneel2026checkerboard,wang2024realization, gong2021quantum, gong2021experimental,zhao2025microwave, braumuller2021probing, yan2019strongly}. Anyonic statistics have only been simulated digitally in superconducting architectures~\cite{xu2023digital,andersen2023non-abelian}.

In this work, we show that superconducting circuits based on transmon qubits can serve as analog simulators of the BH and AH models, and we present an explicit protocol for their implementation. To our knowledge, this is the first proposal for analog simulation of the AH model on a superconducting platform. For the BH model, previous transmon realizations of parametrically tunable interactions were restricted to two particles in total~\cite{wang2026observing}, while our protocol allows up to three particles per site, with no limitation on the total number of particles.
Both results follow from the same mechanism. We find that a sinusoidal modulation of the transmon frequencies, applied in an alternating pattern across the grid, resonantly drives multiple transitions between many-body states. We can control the individual ratios between the transition amplitudes and match them to the hopping amplitudes of the target models. This scheme requires restricting the system to at most three excitations per transmon for the BH model, and two for the AH model. 
Furthermore, we can tune the effective on-site interaction by varying the detuning between the drive frequencies and the transition energy to the higher excited states. Finally, we can engineer complex density-dependent transition amplitudes with the drive phases and simulate a system of interacting anyons. The flexibility of our scheme allows us to simultaneously tune the on-site energy, on-site interaction, and different density-dependent hopping functions, including the statistical anyonic phase.

The paper is structured as follows: in Sec.~\ref{sec:setup} we show the setup of our platform. We present the target BH and AH models and discuss their motivation in Sec.~\ref{sec:target}, and show how to drive the transmon platform to realize the models in Sec.~\ref{Sec:method}. We then simulate the resulting dynamics of our method and compare to those of the exact models in Sec.~\ref{sec:numerics}, and finally discuss the results in Sec.~\ref{sec:conclusion}. 

\section{Setup}
\label{sec:setup}

In this section we describe the experimental platform and the approximations made to reach the effective description of the system.

\begin{figure}[tbp]
  \centering
  \def\svgwidth{0.9\columnwidth}
\begingroup%
  \makeatletter%
  \providecommand\color[2][]{%
    \errmessage{(Inkscape) Color is used for the text in Inkscape, but the package 'color.sty' is not loaded}%
    \renewcommand\color[2][]{}%
  }%
  \providecommand\transparent[1]{%
    \errmessage{(Inkscape) Transparency is used (non-zero) for the text in Inkscape, but the package 'transparent.sty' is not loaded}%
    \renewcommand\transparent[1]{}%
  }%
  \providecommand\rotatebox[2]{#2}%
  \newcommand*\fsize{\dimexpr\f@size pt\relax}%
  \newcommand*\lineheight[1]{\fontsize{\fsize}{#1\fsize}\selectfont}%
  \ifx\svgwidth\undefined%
    \setlength{\unitlength}{94.89543915bp}%
    \ifx\svgscale\undefined%
      \relax%
    \else%
      \setlength{\unitlength}{\unitlength * \real{\svgscale}}%
    \fi%
  \else%
    \setlength{\unitlength}{\svgwidth}%
  \fi%
  \global\let\svgwidth\undefined%
  \global\let\svgscale\undefined%
  \makeatother%
  \begin{picture}(1,0.52776296)%
    \lineheight{1}%
    \setlength\tabcolsep{0pt}%
    \put(0,0){\includegraphics[width=\unitlength,page=1]{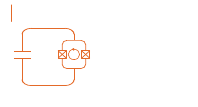}}%
    \put(0.3735586,0.24795981){\color[rgb]{0.89803922,0.41960784,0.16078431}\makebox(0,0)[t]{\lineheight{1.25}\smash{\begin{tabular}[t]{c}\textit{$\Phi$}\end{tabular}}}}%
    \put(0,0){\includegraphics[width=\unitlength,page=2]{qas_schematic_transmononly.pdf}}%
    \put(0.84392183,0.24768578){\color[rgb]{0.11764706,0.18823529,0.72156863}\makebox(0,0)[t]{\lineheight{1.25}\smash{\begin{tabular}[t]{c}\textit{$\Phi$}\end{tabular}}}}%
    \put(0,0){\includegraphics[width=\unitlength,page=3]{qas_schematic_transmononly.pdf}}%
  \end{picture}%
\endgroup%

  \caption{Lumped circuit representation of two capacitively coupled flux-tunable transmons. The frequency of the qubit $\omega_i(\Phi(t))$ depends on the magnetic flux threading the loop of two Josephson junctions, represented by a crossed square.}
  \label{fig:schematictrans}
\end{figure}

For our proposal we assume a lattice of capacitively coupled transmon qubits. The transmon is a type of superconducting qubit consisting of a large capacitance, associated with the energy $E_C$, shunting a Josephson junction, associated with the energy $E_J$, in the regime $E_J/E_C\gg 1$~\cite{koch2007charge-insensitive}. See Fig.~\ref{fig:schematictrans} for the lumped circuit representation of two capacitively coupled transmons, where the crossed square is a Josephson junction. The transmon can be described as an anharmonic oscillator using the bosonic creatoin and annihilation operators $\ad_i$, $\a_i$, and a lattice of capacitively coupled transmons is governed by the Hamiltonian ($\hbar=1$)
\begin{equation}
\label{eq:SC}
    \hat{H}_\text{SC} = \sum_i\omega_i \ad_i \a_i+\frac{\alpha_i}{2}\ad_i\ad_i \a_i \a_i+\sum_{\langle i,j\rangle}g_{ij}\ad
    _i\a_j+h.c.,
\end{equation} 
where $\omega_i=\omega_i^{0,1}$ is the frequency, $\alpha_i=\omega_i^{1,2}-\omega_i^{0,1}$ is the anharmonicity, with $\omega_i^{a,b}$ the energy difference between the transmon energy levels $a$ and $b$, $g_{ij}$ is the hopping amplitude, and $\langle i,j\rangle$ are nearest neighbors. Typically, $\omega_i \sim 5$~GHz, $\alpha_i\sim -250$~MHz, and $g_{ij}\sim 10-100$~MHz. In Eq.~(\ref{eq:SC}) we have omitted the non-particle-conserving terms, such as $\ad_i\ad_j$, which is valid under the rotating wave approximation (RWA) when $|\omega_i+\omega_j|\gg |\omega_i-\omega_j|,|g_{ij}|$. Strictly, since the system is driven and later treated in a rotated frame, these terms should be retained until the RWA is performed in that frame. Here we consider the argument based on the static spectrum of the system for dropping the counter-rotating terms. However, for the numerical simulations in Sec.~\ref{sec:numerics} we keep the counter-rotating terms and confirm that the approximation holds in the regimes we consider. The approximations underlying our effective description are summarized in Table~\ref{tab:approximations}.

When idle, a quantum computer requires a large frequency detuning $\Delta\omega=|\omega_i-\omega_j|$ compared to the hopping amplitude $g_{ij}$ to avoid unwanted cross-talk. However, to perform many-body and entangling operations, qubits need to be able to resonantly interact with each other on demand, which can be achieved by having dynamically tunable transmon frequencies $\omega_i(t)$. Most commonly, the tunable transmon is realized using a SQUID, a closed loop with two Josephson junctions. Then the transmon has a frequency that depends on the local magnetic field threading the SQUID loop, $\omega_i(\Phi(t))=\omega_i^{\text{max}}\sqrt{|\cos(\pi\Phi(t)/\Phi_0)|}[1+d_i^2\tan^2(\pi\Phi(t)/\Phi_0)]^{1/4}$, where $d_i$ is the junction asymmetry and $\Phi_0$ is the magnetic flux quantum~\cite{koch2007charge-insensitive, krantz2019a-quantum}. For a harmonic modulation of the flux $\Phi(t) = \Phi_\text{DC}+\delta_\Phi\cos(\Omega_i t+\varphi_\Phi)$, the transmon frequency $\omega_i(t)=\omega_i[\Phi(t)]$ will be periodic with the same period and can be written as a Fourier series $\omega_i(t)=\omega_i^0+\sum_k\delta_i^k\cos(k\Omega_i t+\varphi_i^k)$~\cite{didier2018analytical}. 

In this paper, we keep only the first-order Fourier expansion, $\omega_i(t)=\omega_i^0+\delta_i\cos(\Omega_i t+\varphi_i)$, which is valid for sufficiently small flux amplitudes $\delta_\Phi$ and a bias point $\Phi_\text{DC}$ that is away from the extrema of $\omega_i(\Phi)$. If $\delta_\Phi$, and consequently $\delta_i$, is too large, the single-mode harmonic approximation breaks down. Instead, one could engineer a flux waveform to compensate for the curvature $\partial^2 \omega_i/\partial\Phi^2$, which gives rise to the higher harmonics $k\geq 2$. We will assume that $\omega_i(t)=\omega_i^0+\delta_i\cos(\Omega_i t+\varphi_i)$ is realizable, with the ability to target specific values for the parameters $\delta_i$, $\Omega_i$, and $\varphi_i$.

\section{Target models}
\label{sec:target}

Here, we describe a variety of interacting bosonic and anyonic lattice models that can be simulated with our quantum analog simulation protocol. The detailed explanation of how to simulate the models with a lattice of flux-tunable transmons is given in Sec.~\ref{Sec:method}.

\subsection{Bose-Hubbard model}
\label{sec:BHM}
The BH model is the canonical description of interacting bosons on a lattice, given by 
\begin{equation}
\label{eq:BH}
    \hat{H}_\text{BH}=\sum_i\mu\ad_i\a_i+\frac{U}{2}\ad_i\ad_i\a_i\a_i+J\sum_{\langle i,j\rangle}\ad_i\a_j+h.c.,
\end{equation}
where $\langle i,j\rangle$ are nearest neighbors.
The many-body physics is governed by two competing energy scales, the on-site interaction $U$, which localizes particles and is associated with potential energy, and the hopping $J$, which delocalizes particles and is associated with kinetic energy. The on-site energy, or chemical potential, $\mu$, sets the filling level of the system.

The BH model is a relatively simple model, but it captures the quantum phase transition from a Mott insulator, a material with a gapped energy spectrum and zero compressibility, to a superfluid, a material in which dissipationless transport of particles can occur. It is a widely studied system, both theoretically and experimentally. The similarity between Eqs.~(\ref{eq:SC}) and~(\ref{eq:BH}) makes the BH model a suitable target model to simulate with transmons. Due to the large anharmonicity, $\alpha$, of the transmon, which is associated with a strong interaction strength $U$, the simulations are often reduced to the hard-core regime~\cite{rosen2025flat-band, rosen2024a-synthetic, guo2021stark, xu2018emulating, praneel2026checkerboard,wang2024realization, gong2021quantum, gong2021experimental,zhao2025microwave, braumuller2021probing, yan2019strongly}, where each site can only hold one particle, i.e., $(\ad_i)^2=0$. However, it is also possible to simulate many-body physics with fixed strong interaction strengths~\cite{mansikkamaki2022beyond, ticea2025observation} and with parametrically tunable interaction strengths, valid only within a two-particle quantum walk~\cite{wang2026observing}. 

Moreover, the BH model admits several extensions that give rise to new physical phenomena. Examples that could be targeted with our protocol include the disordered and tilted BH models, as well as density-dependent tunneling processes. 

For instance, disorder can be introduced through site-dependent on-site energies, $\mu_i$, which allows for the study of Anderson localization in the non-interacting ($U=0$) regime, and many-body localization in the interacting regime. Disordered BH systems have previously been simulated with transmons in the hard-core limit in 1D~\cite{rosen2025flat-band}, and with fixed interactions in 2D~\cite{ticea2025observation,geissler2020mobility}. The disordered BH model is also known to host a third phase, the Bose glass~\cite{gurarie2009phase}.

By adding a linear tilt to the on-site energy, $\mu_i=Fi$, where $F$ is a constant tilt strength, we obtain the 1D tilted BH model, which describes bosonic particles in the presence of a static external field. The tilted BH model can become Stark many-body localized~\cite{yao2020many-body}, which was experimentally observed with transmons in the hard-core limit~\cite{guo2021stark}. The tilted BH model is also known to have chaotic phases~\cite{martin-clavero2025characterization}.

When the hopping amplitude between two sites has a Peierls phase, $Je^{i\theta}$, it is possible to simulate a boson in a magnetic field, and observe Aharonov-Bohm caging when $\theta=\pi$, as demonstrated with transmons in the hard-core boson limit in 2D~\cite{rosen2024a-synthetic} and with a fixed interaction strength in a virtual 3D diamond~\cite{zhang2025synthetic}. Moreover, by adding operators to the Hamiltonian in Eq.~(\ref{eq:BH}) we find the family of extended BH models, such as those with density-dependent tunneling, $J\ad_iF(\n_i,\n_j)\a_j$, also called correlated tunneling. In particular, when $F(\n_i,\n_j)=|\n_i-\n_j|$ a novel pair superfluid phase appears~\cite{rapp2012ultracold}.

\subsection{Anyon-Hubbard model}
\label{sec:AHM}
When two indistinguishable particles are exchanged, the phase difference between the initial and resulting states can either be 0, if the particles are bosons, or $\pi$, if they are fermions. However, in 2D or lower, there also exist quasi-particles called Abelian anyons, which can have an arbitrary phase difference $\theta$ between the exchanged states. In contrast, non-Abelian anyons undergo more general unitary transformations. In this work, we focus on the Abelian case, for which the commutation relations of the anyonic annihilation and creation operators are
\begin{align*}
    \hat{c}_j\hat{c}_k^\dagger-e^{-i\theta \text{sgn}(j-k)}\hat{c}_k^\dagger \hat{c}_j&=\delta_{jk},\\
    \hat{c}_j\hat{c}_k-e^{-i\theta\text{sgn}(j-k)}\hat{c}_k\hat{c}_j&=0,
\end{align*}
where $\text{sgn}(x)$ is the sign function. Interacting anyons on a tight-binding lattice are captured by the AH model, which in 1D can be described using bosonic operators and a density-dependent Peierls phase~\cite{keilmann2011statistically, greschner2015anyon} as
\begin{equation}\label{eq:AHM}
    \hat{H}_{AH}=\sum_i\mu\ad_i \a_i+ \frac{U}{2}\a^\dagger_i\a_i^\dagger \a_i \a_i +J\sum_{i} \left(\a_i^\dagger \a_{i+1}e^{i\n_i\theta}+h.c.\right).
\end{equation}

Shortly after the quasi-particle was proposed, the excitations of fractional quantum Hall states were recognized as anyons~\cite{haldane1991fractional,halperin1984statistics}. Since then, many properties of interacting anyons have been established. The AH model undergoes the Mott-insulator-to-superfluid phase transition by changing only the statistics, i.e., varying $\theta$~\cite{keilmann2011statistically}. Another characteristic of the AH model is that the fractional statistics of anyons leads to asymmetric quantum walks~\cite{liu2018asymmetric}. Even for non-interacting anyons ($U=0$), two anyons in the standard AH model form novel bound states~\cite{kwan2024realization}. Additionally, it is straightforward to extend the AH model to include tilted or disordered on-site energies, similarly to how such terms are incorporated in the BH model.
Although beyond the scope of our protocol, previous work has extended the AH model in 1D lattices to include nearest-neighbor interactions $\n_i\n_j$, leading to additional phase transitions in $\theta$, such as the Haldane insulator~\cite{lange2017anyonic}.

\section{Method}\label{Sec:method}
In this section, we explain how to simulate the BH and AH models described in Sec.~\ref{sec:target}. By driving the transmons' frequencies in an alternating manner, we obtain an effective Hamiltonian matches the target system in a restricted excitation manifold. The approximations underlying this effective description, and the regimes in which they hold, are collected in Table~\ref{tab:approximations}. Our protocol allows the individual on-site energies $\mu_i$ to be tuned directly through the static energies $\omega_i^0$. For systems with at most three particles per site, the on-site interaction $U$ can be tuned by the drive frequencies $\Omega_i$. We can also implement correlated hopping functions $F(\n_i-\n_j)$, satisfying $F(x)=F(-x)$, by changing the ratio between the modulation amplitude and frequency, $\lambda_i=\delta_i/\Omega_i$. Furthermore, for systems with at most two particles per site, we find that we can simulate the 1D AH model with tunable $\mu_i$, $U$, and statistical phase $\theta$ by controlling the drive phases $\varphi_i$.

\subsection{Frequency modulation of transmons}
We consider the rotated frame $\hat{H}'=\hat{U}\hat{H}\hat{U}^\dagger +i\dot{\hat{U}}\hat{U}^\dagger$ of Eq.~(\ref{eq:SC}), with 
\begin{equation}\label{eq:BHrotframe}
    \hat{U}=\prod_j\exp\biggl(i\int_0^td\tau (\omega_j(t)-\omega_j^0)\ad_j \a_j +\frac{h_j}{2}\ad_j \ad_j \a_j \a_j\biggr),
\end{equation} 
and recall that the longitudinally driven transmon energies are given by $\omega_i(t)=\omega_i^0+\delta_i\cos(\Omega_i t+\varphi_i)$. The frame serves two purposes, it makes the diagonal energies time-independent and sets an effective anharmonicity through the frame parameter $h_j$. We see this when we consider the effect of changing the frame on the diagonal part of Eq.~(\ref{eq:SC}), $\hat{H}_0=\sum\omega_i\ad_i\a_i+\alpha_i/2\ad_i\ad_i\a_i\a_i$, which gives
\begin{equation}
\hat{H}_0'=\sum_i\omega_i^{0}\ad_i \a_i+\frac{\alpha_i-h_i}{2}\ad_i\ad_i\a_i\a_i.
\end{equation}
In this frame the qubit frequency is $\omega_i^0$ and the effective anharmonicity is $\alpha_i-h_i$. To match the on-site energy and on-site interaction of the models described in Sec.~\ref{sec:target} we simply set $\omega_i^0$ through the DC bias $\Phi_\text{DC}$, and $U=\alpha_j-h_j$ by choosing frame parameter $h_j$. However, $h_j$ also enters the off-diagonal Hamiltonian, changing the conditions for resonant transitions between neighboring transmon states. To see this, we take the hopping part of Eq.~(\ref{eq:SC}), $\hat{H}_I = \sum_{\langle i,j\rangle}g_{ij}(\ad_i\a_j+h.c.)$, which transforms into
\begin{equation}
\label{Eq:HIprim}
\begin{split}
\hat{H}_I'&=\sum_{\langle i,j\rangle}\sum_{n_i,n_j=0}^\infty\sum_{k_i,k_j=-\infty}^{\infty}\sqrt{(n_i+1)(n_j+1)}  \\
&\times g_{ij}J_{k_i}(\lambda_i)J_{k_j}(\lambda_j)e^{i(h_in_i-h_jn_j+k_i\Omega_i-k_j\Omega_j)t}\\
&\times e^{i(k_i\varphi_i-k_j\varphi_j)}\ket{n_i+1,n_j}\bra{n_i,n_j+1}+h.c.,\end{split}
\end{equation}
where we have written $\ad_i a_j$ in the Fock-basis, and used the Jacobi-Anger expansion 
\begin{equation}
    e^{i\lambda_i\sin(\Omega_i t +\phi_i)}=\sum_{k_i=-\infty}^\infty J_{k_i}(\lambda_i)e^{ik_i(\Omega_i t+\phi_i)},
\end{equation}
where $J_{k_i}(\lambda_i)$ is the Bessel function of the first kind with order $k_i$, and the argument $\lambda_i=\delta_i/\Omega_i$. 

\begin{figure}[hbtp]
  \centering
  \def\svgwidth{\columnwidth}
  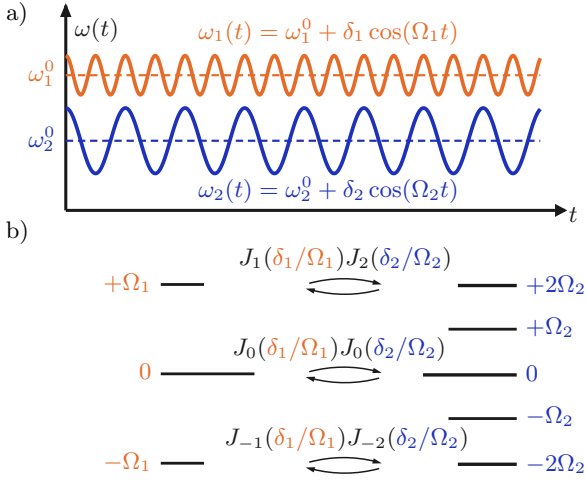
  \caption{Sideband diagram of frequency-modulated transmons. a) The time-dependent frequencies $\omega_i(t)=\omega_i^0+\delta_i\cos(\Omega_i t)$ ($i=1,2$) create sidebands. b) Sidebands associated with the $\ket{10}\bra{01}$ transition in Eq.~(\ref{Eq:HIprim}) have frequencies $k\Omega_i$, with integer $k$. The length of each line is proportional to the sideband amplitude, given by the Bessel functions $J_k(\delta_i/\Omega_i)$. In this example, resonant exchange occurs when sidebands from the two transmons have the same frequency, with an effective coupling $J_{k_1}(\delta_1/\Omega_1)J_{k_2}(\delta_2/\Omega_2)$. The total transition amplitude is obtained by summing over all resonant sideband pairs, $J=J_1J_2 + J_0 J_0 + J_{-1}J_{-2}$.}
  \label{fig:schematicsideband}
\end{figure}

In this rotated frame, the $k_i\Omega_i$ terms in the exponent of Eq.~(\ref{Eq:HIprim}) can be interpreted as sidebands at integer multiples of the drive frequencies $\Omega_i$, with sideband amplitude given by the Bessel function of that integer order~\cite{silveri2017quantum}, see Fig.~\ref{fig:schematicsideband}. For a transition $\ket{n_i+1,n_j}\bra{n_i,n_j+1}$ to occur resonantly, the energy difference between the sidebands $k_i\Omega_i-k_j\Omega_j$ needs to cancel the energy $n_ih_i-n_jh_j$, making the time-dependent exponent in Eq.~(\ref{Eq:HIprim}) zero. For example, the transition $\ket{10}\bra{01}$ has $n_ih_i-n_jh_j=0$, so it is resonant for $(k_i,k_j)=(0,0)$, independent of the drive frequency $\Omega_i$. Because transitions are associated with different energies $h_in_i-h_jn_j$, they have different resonance conditions, e.g., the transition $\ket{20}\bra{11}$ has $n_ih_i-n_jh_j=h_i$, so it is not in resonance for the pair of Bessel orders $(k_i,k_j)=(0,0)$.

Furthermore, when both transmons are driven, there exist infinitely many combinations of Bessel orders $k_i$ and $k_j$ that satisfy the resonance condition for every transition, and each pair has the reduced transition amplitude $J_{k_i}(\lambda_i)J_{k_j}(\lambda_j)$. 

\subsection{Resonance conditions for Fock-state transitions}
To simulate a model, the transition amplitudes in Eq.~(\ref{Eq:HIprim}) must match the corresponding amplitudes of the target models, so the sideband amplitudes for all relevant transitions need to be tuned simultaneously. The remainder of Sec.~\ref{Sec:method} studies how to do this.

Now, we will assume that all transmons have equal anharmonicity, $\alpha_i=\alpha$, as we want equal effective anharmonicity we set $h_i=h$. The resonance condition for the $\ket{n_i+1,n_j}\bra{n_i,n_j+1}$ transition, taken from the time-dependent exponent in Eq.~(\ref{Eq:HIprim}), becomes
\begin{equation}\label{eq:resonance}
    k_i\Omega_i-k_j\Omega_j+mh=0,
\end{equation}
where $m=n_i-n_j$. If we consider $\nmax$ as the maximum number of particles allowed per site, then $n_i,n_j\leq\nmax-1$, and $m$ takes $2\nmax-1$ integer values, ranging from $-\nmax +1$ to $\nmax-1$. We note that different Fock-state transitions can give the same value for $m$, e.g., $\ket{10}\bra{01}$ and $\ket{21}\bra{12}$ both have $m=0$. As Eq.~(\ref{eq:resonance}) must be satisfied for multiple $m$ using the two drive frequencies, $\Omega_i$ and $\Omega_j$, we consider frequencies of the form $\Omega_i=ph/r$ and $\Omega_j=qh/r$, where $p,q,r$ are integers. 

Now, let $K^m$ be the set of Bessel order pairs, $(k_i,k_j)$, that satisfy Eq.~(\ref{eq:resonance}) for a given $m$. As an example, if $\Omega_i=3h/2$, $\Omega_j=h/2$, the $m=0$ transition $\ket{10}\bra{01}$ is resonant for the pairs of Bessel orders $(0,0)$, $(1,3)$, and $(-1,-3)$, so they are in $K^0$, while $(-1,-1)$ and $(0,2)$ are in $K^1$, associated with the transition $\ket{20}\bra{11}$. Note that the $K^m$ sets contain infinitely many elements. As $J_{k_i}(z)\ll 1$ for small $z$ and large $|k_i|$, we only consider only a finite subset, $|k_i|\leq10$.

The pairs of Bessel orders not in $K^m$ are part of terms in Eq.~(\ref{Eq:HIprim}) where the exponent is not zero, and become time-dependent hopping amplitudes with an oscillating phase. Following the previous example, we find that $(-1,-1)\notin K^0$, so for the transition $\ket{10}\bra{01}$ the pair becomes the term $J_{-1}(\lambda_i)J_{-1}(\lambda_j)e^{-iht}\ket{10}\bra{01}$. We assume that these oscillating amplitudes can be removed with an RWA, which requires the oscillation frequency $k_i\Omega_i-k_j\Omega_j+mh$, with $(k_i,k_j)\notin K^m$, to be much larger than the associated transition amplitude $g_{ij}J_{k_i}(\lambda_i)J_{k_j}(\lambda_j)$. We find it useful to consider the worst-case ratio
\begin{equation}\label{eq:rmin}
   r_\text{min} =\min_{(k_i,k_j)\notin K^m}\left[\frac{|k_i\Omega_i-k_j\Omega_j+mh|}{|g_{ij}J_{k_i}(\lambda_i)J_{k_j}(\lambda_j)|}\right],
\end{equation} and use $r_\text{min}$ to guide the choice of $\Omega_i$ and $\lambda_i$. See Table~\ref{tab:approximations} for a summary of approximations and relevant parameter regimes. 

We can now define the effective transition amplitude for each transition with the same $m=n_i-n_j$ value as 
\begin{equation}\label{eq:Gm}
    G^m(\lambda_i,\lambda_j)=g_{ij}\sum_{\mathclap{(k_i,k_j)\in K^m}}J_{k_i}(\lambda_i)J_{k_j}(\lambda_j)e^{i(k_i\varphi_i-k_j\varphi_j)}.
\end{equation}

So far we have only considered one pair of transmons $i,j$. However, for our target models we need the transition amplitudes for an $m$-transition to have the same value for all coupled transmons in the lattice. The simplest way to do this is to drive the transmons in an alternating manner, creating two sublattices of drive parameters.

The next step is to tune the transition amplitudes $G^m$ by numerically finding drive parameters that match the transition amplitudes of a target model, e.g., the constant $G^m=G$ for the BH model. 
Following the truncation introduced previously for Eq.~(\ref{eq:resonance}), we allow at most $\nmax$ excitations per site, such that $(\ad)^{\nmax+1}=0$. Then, there are $2\nmax-1$ different $m$ values, each associated with its own $G^m$ that we need to tune. Given our limited number of parameters for tuning $G^m$, it is beneficial to reduce the number of distinct values, which we do in Sec.~\ref{sec:symmetries} by considering the relationship between the $\pm m$ transitions.

\subsection{Bose-Hubbard model and correlated hopping through symmetric transitions}
\label{sec:symmetries}
In this section we assume that we are driving without phases, $\omega_i(t)=\omega_i^0+\delta_i\cos(\Omega_it)$. 
If a Bessel pair $(k_i,k_j)$ satisfies Eq.~(\ref{eq:resonance}) for $m=n_i-n_j$ and lies in $K^m$, then multiplying Eq.~(\ref{eq:resonance}) by $-1$ places $(-k_i,-k_j)$ in $K^{-m}$, so the $\pm m$ transitions share the same Bessel orders but with opposite signs. 

We return to the example of $(\Omega_i,\Omega_j)=(3h/2,h/2)$, where the $m=1$ transition $\ket{20}\bra{11}$ is resonant for the Bessel pair $(-1,-1)\in K^1$. Swapping the signs of the Bessel orders gives $3h/2-h/2=h$, so the $m=-1$ transition $\ket{11}\bra{02}$ transition is resonant for $(1,1)\in K^{-1}$. For the $m=0$ transitions, there is no corresponding $-m$ transition, and the negative and positive versions of each Bessel order pair satisfy the resonance condition. For our example both $(1,3)$ and $(-1,-3)$ are then in $K^0$.

Using $J_{-k}(z)=(-1)^kJ_k(z)$, the shared structure of $K^m$ and $K^{-m}$ lets us write the transition amplitudes of the $\pm m$ transitions with the same sum, such that
\begin{equation}
    G^m-G^{-m}=g_{ij}\sum_{K^m} J_{k_i}(\lambda_i)J_{k_j}(\lambda_j)(1-(-1)^{k_i+k_j}).
\end{equation}
Now, we can choose the frequencies $\Omega_i$ such that the sums $k_i+k_j$, for every pair $(k_i,k_j)\in K^m$, have equal parity, and $(-1)^{k_i+k_j}$ is either always positive or always negative, meaning that
\begin{equation}\label{eq:Gmink}
    G^m=(-1)^{k_i+k_j}G^{-m}.
\end{equation} 
Crucially, the magnitudes are equal, $|G^m|=|G^{-m}|$, reducing the number of unique $|G^m|$ to only $\nmax$. 

\begin{table*}[htbp]
\caption{\label{tab:approximations} Approximations underlying the effective description, the regime in which each holds, and its role.}
\begin{tabular}{m{0.25\textwidth}m{0.33\textwidth}m{0.33\textwidth}}
\hline\hline
        \textbf{Approximation} & \textbf{Regime} & \textbf{Role} \\ \hline
        
        Rotating-wave approximation
            & $|\omega_i+\omega_j|\gg|\omega_i-\omega_j|,\,|g_{ij}|$
            & Reduces the coupled-transmon Hamiltonian to the particle-conserving form, Eq.~(\ref{eq:SC}). \\\hline
        
        Linearized flux modulation
            & $\Phi_\text{DC}$ away from extrema of $\omega_i(\Phi)$, or \newline precise control of non-harmonic $\Phi(t)$
            & Single-tone frequency modulation $\omega_i(t)=\omega_i^0+\delta_i\cos(\Omega_i t+\varphi_i)$. \\\hline
        
        Uniform anharmonicities \newline $\alpha_i=\alpha$
            & 
            & Allows for the same resonance condition from Eq.~(\ref{eq:resonance}) for any pair of transmons across the array. \\\hline
        
        Rotating-wave approximation \newline in the rotating frame

            & $\frac{|k_i\Omega_i-k_j\Omega_j+mh|}{ |g_{ij}J_{k_i}(\lambda_i)J_{k_j}(\lambda_j)|}\gg1$ for $(k_i,k_j)\notin K^m$
            & Removes the sidebands not satisfying the resonance condition, such that the transition amplitude is given by $G^m$ of Eq.~(\ref{eq:Gm}) \\\hline
        
        Removing higher order Bessel \newline functions &  $|k_i|\leq 10$  & Enables numerical solutions for Eqs.~(\ref{eq:Gm}) and~(\ref{eq:argg1}).\\\hline
        
        Linear amplitude of $\omega_i(t)$ &$\delta_i\lesssim 1.5$ GHz  & Allows $\alpha p\lambda_i/r, \alpha q\lambda_j/r\lesssim 7.5$ to get a larger search space for simulation parameters.\\\hline\hline
\end{tabular}
\end{table*}
Moreover, when all $k_i+k_j$ are even the amplitudes are equal, $G^m=G^{-m}$, while all-odd $k_i+k_j$ give $G^m=-G^{-m}$, i.e., a $\pi$ phase between the $m$ and $-m$ transition amplitudes. In Appendix~\ref{app:phase} we show that for some configurations of total particles $N_\text{total}=\sum_i\langle \n_i\rangle $, and maximum particles per site, $\nmax$, the $\pi$ phase can be absorbed into a frame that leaves diagonal observables unchanged. However, for most configurations, the $\pi$ phase changes the dynamics, and models such as the BH are not simulable. To avoid the phase, we need to choose drive frequencies that result in even $k_i+k_j$, e.g., $\Omega_i=3h/2$, $\Omega_j=h/2$. The choice of only driving every other transmon, $\Omega_i= h$, $\Omega_j=0$ instead gives $(-1,0)\in K^1$, with $-1+0$ odd. This is the regime of previous work on tunable interactions in Ref.~\cite{wang2026observing}, a regime that we find limits the entire system to two particles, because the $\pi$ phase between the $G^1$ and $G^{-1}$ amplitudes cannot be resolved for systems with three or more particles in total.

With the $\pi$ phase removed, the individual transition amplitudes $G^{|m|}$ can be tuned independently. This freedom admits a natural interpretation as a density-dependent hopping, $J[\ad_iF(\n_i-\n_j)\a_j+h.c.]$, where $JF(\n_i-\n_j)=G^{|m|}$, and $F$ is an even function. Fixing $F(0)=1$ identifies the hopping amplitude as $J=G^{0}$, and the density-dependent function as $F(\n_i-\n_j)=G^{|m|}/G^0$. For three excitations per site, we only need to solve for $|m|=1,2$, so implementing a target $F$ reduces to matching the two values $F(1)$ and $F(2)$. For example, the BH model has equal hopping amplitude, $F(1)=F(2)=1$, while the correlated hoppings studied in Refs.~\cite{rapp2012ultracold, meinert2016floquet} have $F(x)=J_0(x)$, the zeroth-order Bessel function. Given the target values, we numerically solve for the $(\lambda_i,\lambda_j)$ that satisfy the two independent transcendental equations
\begin{equation}\label{eq:c_mGm}
    G^{|m|}(\lambda_i,\lambda_j)-F(|m|)G^{0}(\lambda_i,\lambda_j)=0,
\end{equation}
for $|m|=1,2$.
It is important to remember that the $\lambda_i$ are the experimental parameters $\delta_i/\Omega_i$. Assuming $\delta_i\lesssim1.5$ GHz, and $\Omega_i=2.5\alpha\sim-600$ MHz we get $\lambda_i\lesssim 2.5$. See Table~\ref{tab:approximations} for a summary of approximations.

To summarize, we are able to realize the Hamiltonian
\begin{equation}\begin{split}
    H_\text{eff}&=\sum_i\mu_i\ad_i a_i+\frac{U}{2}\ad_i\ad_i\a_i\a_i\\
    &+J\sum_{\langle i,j\rangle}\ad_iF(\n_i-\n_j)\a_j+h.c.,
    \end{split}
\end{equation}
where $(\ad_i)^4=0$. The $\mu_i$, $U$, and even function $F(\n_i-\n_j)$ are parametrically tunable. The $\mu_i$ is set directly by the static part of the transmon energy, $\omega_i^0$, while the interaction strength $U=\alpha-h$, is determined by the frame parameter $h$, which relates to the drive frequencies $\Omega_i={ph}/{r}$, $\Omega_j={qh}/{r}$, where we are free to choose $p$, $q$, and $r$ within the constraints described in this section. The even function $F(\n_i-\n_j)$ is realized by finding the $(\lambda_i,\lambda_j)$ that solve Eq.~(\ref{eq:c_mGm}) for $|m|=1,2$, where $m=n_i-n_j$. 

\subsection{Anyon-Hubbard model through drive phases}\label{sec:phase}
Now we add phases $\varphi_i$ to the transmon frequency modulation, so that $\omega_i(t)=\omega_i^0+\delta_i\cos(\Omega_i t+\varphi_i)$, which in practice can be done by applying a phase-offsetting the flux drives. Each sideband then acquires a phase determined by its Bessel orders, $\exp[i(k_i\varphi_i - k_j\varphi_j)]$, so the terms in Eq.~(\ref{eq:Gm}) can interfere destructively, modifying both the magnitude and the phase of $G^m$. In this section we determine the effective phase of each transition amplitude, while ensuring that the magnitudes are equal across an array, with the goal of simulating the AH model.

Recalling that the drive frequencies are given by $\Omega_i=ph/r$, $\Omega_j=qh/r$, we find that the Bessel orders in $K^m$ can be written in terms of any base pair $(k_i^0,k_j^0)$ and the strides $p$ and $q$, as $(k_i^s,k_j^s)=(k_i^0,k_j^0)+(q,p)s$, $s\in\mathbb{Z}$. We return to our example of $(\Omega_i,\Omega_j)=(3h/2,h/2)$, where $(0,2)$, $(-1,-1)$, and $(-2,-4)$ all satisfy $k_i\Omega_i-k_j\Omega_j+h=0$ and are in $K^1$. Letting $(-1,-1)=(k_i^0,k_j^0)$, the other two pairs are $(-1,-1)\pm(1,3)$. Returning to the phase factor, $\exp[i(k_i\varphi_i - k_j\varphi_j)]$, we can similarly decompose each phase into $(k_i^0\varphi_i-k_j^0\varphi_j)+(q\varphi_i-p\varphi_j)s=\beta_{ij}+\gamma_{ij}s$. The phase of the base pair, $\beta_{ij}$, is part of every phase factor in $G^m$ and, as such, does not change the magnitude.

Now, instead of summing over all Bessel orders in $K^m$, we can sum over the integer $s$, and write the transition amplitude as
\begin{equation}
    G^m=g_{ij}e^{i\beta_{ij}}\sum_sJ_{k_i^s}(\lambda_i)J_{k_j^s}(\lambda_j)e^{i\gamma_{ij} s}.
\end{equation}
Since the magnitudes of $G^m$ must be equal across an array, $\gamma_{ij}$ is restricted to two values, $\gamma_{ij}\in\{-\gamma,\gamma\}$, for fixed $\gamma$. When all $\gamma_{ij}$ take the same value, the physical phases of the drives, $\varphi_i$ and $\varphi_j$, must alternate, as the drive frequency and amplitude do. The effective phase, which is given by the argument of the complex transition amplitude $\phi^m_{ij}=\arg(G^m)$, will then be uniform for all transmon pairs. If the transition of every coupled pair of transmons should have different phases, $\gamma_{ij}$ must alternate along the transmon array as $\gamma_{i,j}=\gamma$ and $\gamma_{j,i'}=-\gamma$.

Moreover, from Sec.~\ref{sec:symmetries} we know that all Bessel orders in $K^0$ are either $(0,0)$, which carries no phase, or conjugate pairs $(k_i,k_{j})$ and $(-k_{i}, -k_{j})$ with opposite phase contributions. The effective phase of the $m=0$ transitions therefore vanishes, $\phi_{ij}^0=0$. Similarly, since we know that $(k_i,k_{j})\in K^m$ implies $(-k_{i},-k_{j})\in K^{-m}$, the effective phases on the $\pm m$ transitions have opposite signs, $\phi_{i,j}^m=-\phi_{i,j}^{-m}$. For two particles per site, the hopping operator in the Fock basis is then given by \begin{equation}\label{eq:BHphase}\begin{split}
    g_{ij}\ad_i \a_j&=G^0(\ket{10}\bra{01}+2\ket{21}\bra{12})\\&+\sqrt{2}G^1(\ket{20}\bra{11}e^{i\phi^1_{ij}}+\ket{11}\bra{02}e^{-i\phi^1_{ij}}).
    \end{split}
\end{equation}

Let us now compare Eq.~(\ref{eq:BHphase}) to the transitions of our target AH model. We restrict ourselves to two particles per site, $(\ad_i)^3=0$, as we do not have enough experimental parameters to match the AH model for more particles. We return to this at the end of this section. Additionally, the model is only valid in 1D, so neighboring sites can be indexed with the indices $\ell,\ell+1$. Then, the hopping operator of the AH model from Eq.~(\ref{eq:AHM}) in the Fock basis is
\begin{equation}\label{eq:AHMhopping}\begin{split}
    J\a_\ell^\dagger \a_{\ell+1}e^{in_\ell\theta}&=J\biggl[\ket{10}\bra{01}+\sqrt{2}\ket{20}\bra{11}e^{i\theta}\\
    &+\sqrt{2}\ket{11}\bra{02}+2\ket{21}\bra{12}e^{i\theta}\biggr].
    \end{split}
\end{equation}
\begin{figure}[htbp]
  \centering
  \def\svgwidth{0.85\columnwidth}
  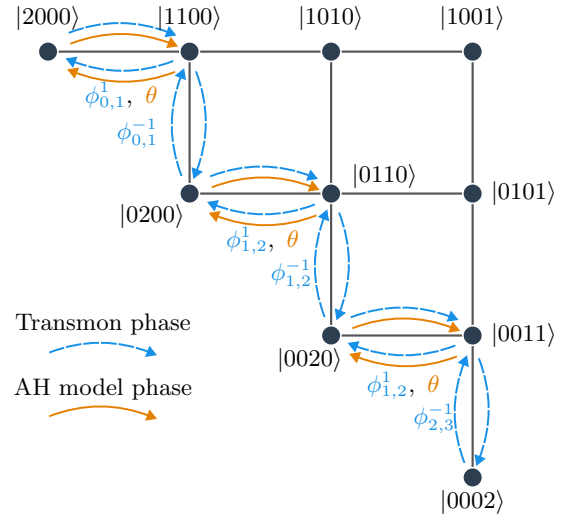
  \caption{A Fock-state lattice representation of four sites with two particles. Each vertex is a many-body Fock state, and each edge represents two states being coupled by $J(\ad_i\a_j+h.c.)$. The curved arrows indicate the phase of the hopping amplitude, where the blue dashed arrows are the hopping phases for the transmon system in Eq.~(\ref{eq:BHphase}) and the orange solid arrows are for the AH model in Eq.~(\ref{eq:AHMhopping}). The transmon phases $\phi_{\ell,\ell+1}^0=0$ are not included in the figure. The total phase around any closed loop is the same for both systems when $\phi_{\ell,\ell+1}^{\pm1}=\pm\ell\theta$.}
  \label{fig:fockstatelattice}
\end{figure}
Next, we introduce an additional rotated frame $U=\prod_\ell e^{-i\phi^1_{\ell-1,\ell}\ket{2}\bra{2}_\ell}$, such that $\ket{2}_\ell\to e^{-i\phi_{\ell-1,\ell}}\ket{2}_\ell$, to identify what phases $\phi_{\ell,\ell+1}^1$ implement the AH model. The transmon hopping operator in Eq.~(\ref{eq:BHphase}) is rotated as
\begin{equation}\label{eq:BHphaserotated}\begin{split}
    g_{ij}\hat{a}_\ell^{\dagger}\a_{\ell+1}&=G^0\biggl[\ket{10}\bra{01}+2\ket{21}\bra{12}e^{i(\phi^1_{\ell,\ell+1}-\phi_{\ell-1,\ell}^1)}\\
    &+\sqrt{2}\ket{20}\bra{11}e^{i(\phi^1_{\ell,\ell+1}-\phi^1_{\ell-1,\ell})}\\
    &+\sqrt{2}\ket{11}\bra{02}e^{-i(\phi^{1}_{\ell,\ell+1}-\phi^1_{\ell,\ell+1})}\biggr],\\
    \end{split}
\end{equation} 
where we assume $G^0=G^1$.
The phase on the $\ket{11}\bra{02}$ transition, $\phi^{1}_{\ell,\ell+1}-\phi^1_{\ell,\ell+1}$, is cancelled by the
frame transformation. If $\phi_{\ell,\ell+1}^1=\ell\theta$, i.e., the
effective phase of the $m=1$ transition ramps linearly along the chain
in steps of the statistical phase $\theta$, then the phases on the
$\ket{21}\bra{12}$ and $\ket{20}\bra{11}$ transitions are
$\theta$, making Eqs.~(\ref{eq:AHMhopping})
and~(\ref{eq:BHphaserotated}) equal. We can visualize this equivalence with a Fock-state lattice, and consider the phases across all closed loops~\cite{dalibard2011colloquium:, zhang2025synthetic}, see Fig.~\ref{fig:fockstatelattice}.

Since each pair of transmons needs a different phase on its $m$-transition, the $\gamma_{\ell,\ell+1}=q\varphi_\ell-p\varphi_{\ell+1}$ must alternate in sign along neighboring pairs of the array, $\gamma_{\ell-1,\ell}=-\gamma_{\ell,\ell+1}$. To achieve this, we 
consider two linear ramps for the phases $\varphi_{2\ell}=\varphi_0-2\ell {\gamma}/{q}$ and $\varphi_{2\ell+1}=\varphi_1-2\ell {\gamma}/{p}$ of the even- and odd-indexed sites, respectively. This choice of driving phases requires that the drive frequency $\Omega_i=ph/r$ is assigned to the even-indexed sites. The initial phases $\varphi_0$ and $\varphi_1$ are fixed by the two boundary constraints $q\varphi_0-p\varphi_1 = \gamma$ and the $m=1$ phase of the first transmon pair $\phi_{0,1}^1=0$. To illustrate, we work out the first four $\gamma_{\ell,\ell+1}$
\begin{align*}
    \gamma_{0,1}&=q\varphi_0-p\varphi_1=\gamma,\\
    \gamma_{1,2}&=q\varphi_0-p\varphi_1-2q\gamma/q=-\gamma,\\
    \gamma_{2,3}&=q\varphi_0-p\varphi_1-2q\gamma/q+2p\gamma/p=\gamma,\\
    \gamma_{3,4}&=q\varphi_0-p\varphi_1-4q\gamma/q+2p\gamma/p=-\gamma.
\end{align*}
\begin{figure*}[htbp]
    \centering
    \includegraphics[width=\textwidth]{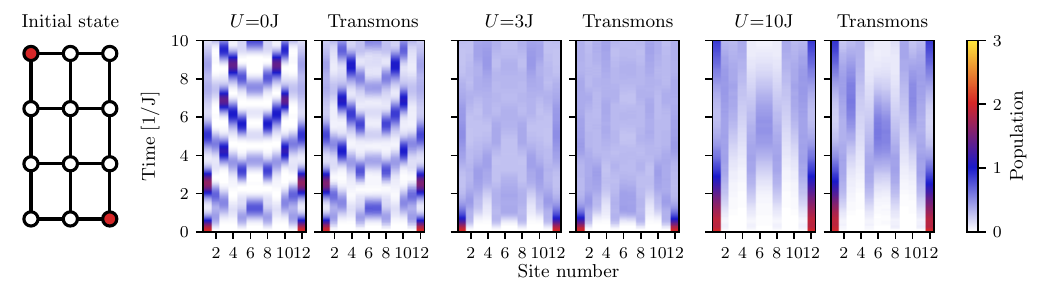}
    \caption{A quantum walk of two particles initialized in each of sites one and twelve in a $4\times3$ grid for $U=\{0,3J,10J\}$. For each interaction strength, the left plot shows the exact dynamics of the BH model, Eq.~(\ref{eq:BH}), and the right plot, the simulated transmon system. The system contains four particles, which exceeds the maximum occupation per site, $\nmax=3$, so the truncation is not guaranteed by particle-number conservation.}
    \label{fig:doublon}
\end{figure*}
\begin{figure*}[htbp]
    \centering
    \includegraphics[width=\textwidth]{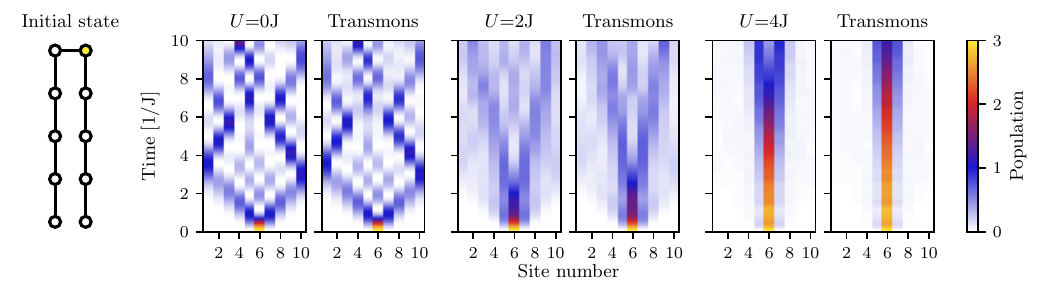}
    \caption{A quantum walk of a triplon initialized at site six in a ten site 1D chain for $U=\{0,2J,4J\}$. For each interaction strength, the left plot shows the exact dynamics of the BH model, Eq.~(\ref{eq:BH}), and the right plot, the simulated transmon system.}
    \label{fig:triplon}
\end{figure*}
From $\phi_{\ell,\ell+1}^1=\ell \theta$, the difference between the $m=1$ phases of consecutive transmon pairs, $\Delta\phi_{\ell,\ell-1}^1=\phi_{\ell,\ell+1}^1-\phi_{\ell-1,\ell}^1$, must be equal to the statistical phase $\theta$ of the AH model. The three sites involved in this difference will either be two sites with even indices and one with odd index, or vice versa, giving the phase differences
\begin{align*}
\Delta \phi_{\ell,\ell-1}^1&=-\frac{2k_i^0\gamma}{q}-2\arg\left(\sum_sJ_{k_i^s}(\lambda_i)J_{k_j^s}(\lambda_j)e^{i\gamma s}\right),\\
\Delta\phi_{\ell,\ell-1}^1&=+\frac{2k_j^0\gamma}{p}+2\arg\left(\sum_sJ_{k_i^s}(\lambda_i)J_{k_j^s}(\lambda_j)e^{i\gamma s}\right),
\end{align*}
respectively. The two are equal when
\begin{equation}\label{eq:argg1}
    \arg\left(\sum_sJ_{k_i^s}(\lambda_i)J_{k_j^s}(\lambda_j)e^{i\gamma s}\right)=-\frac{pk_i^{0}+qk_j^{0}}{2pq}\gamma.
\end{equation} 
Using Eq.~(\ref{eq:argg1}) to calculate $\Delta\phi_{\ell,\ell-1}^1$, we find that both phase differences are equal to $\theta=\gamma {r}/{pq}$, where we used the resonance condition for a base pair in $K^1$, $k_i^0ph/r-k_j^0qh/r+h=0$. Lastly, we numerically solve for the pair of ratios of drive parameters $(\lambda_i,\lambda_j)$ that satisfy Eq.~(\ref{eq:argg1}), to get the linear ramp $\phi^1_{\ell,\ell+1}=\ell\theta$, and equal magnitudes of the hopping amplitudes, $|G^0(\gamma)|=|G^1(\gamma)|$.

We return to the example of $(\Omega_i,\Omega_j)=(3h/2,h/2)$, and show an implementation of $\theta=0.1$ on seven sites in Table~\ref{tab:phasesAHM}.  
\begin{table}[htbp]
\caption{\label{tab:phasesAHM} Example of the AH drive-phase
construction for a seven-site chain with
$(\Omega_i,\Omega_j)=(3h/2,\,h/2)$, and base pair
$(k_i^0,k_j^0)=(-1,-1)\in K^1$. The boundary phases are
$\varphi_0=0.075$, $\varphi_1=-0.025$.}
\begin{tabular}{C{0.2\columnwidth }C{0.2\columnwidth }C{0.2\columnwidth }C{0.15\columnwidth }C{0.15\columnwidth }}
\hline\hline
        $(\ell,\ell+1)$ & $\gamma_{\ell,\ell+1}$ & $-\frac{pk_i^{0}+qk_j^{0}}{2pq}\gamma$ & $\beta_{\ell,\ell+1}$& $\phi^1_{\ell,\ell+1}$ \\ \hline

        $(0,1)$ & $+0.15$ & $+0.1$ & $-0.1$ & $0$    \\

        $(1,2)$ & $-0.15$ & $-0.1$ & $+0.2$ & $0.1$ \\

        $(2,3)$ & $+0.15$ & $+0.1$ & $+0.1$ & $0.2$\\

        $(3,4)$ & $-0.15$ & $-0.1$ & $+0.4$ & $0.3$  \\

        $(4,5)$ & $+0.15$ & $+0.1$ & $+0.3$ & $0.4$  \\

        $(5,6)$ & $-0.15$ & $-0.1$ & $+0.6$ & $0.5$ \\\hline\hline
\end{tabular}
\end{table}

Finally, we consider three particles per site. The drive phases $\varphi_i$ and $\varphi_j$ are already fixed by the requirement to realize the value of $\gamma$ set by the choice of $\theta$, so the remaining conditions must be satisfied with $\lambda_i$ and $\lambda_j$ alone. This leaves three equations, $|G^0|=|G^1|$, $|G^0|=|G^2|$, and Eq.~(\ref{eq:argg1}), for two parameters, and the system becomes over-constrained.
\begin{figure*}[htbp]
    \centering
    \includegraphics[width=\textwidth]{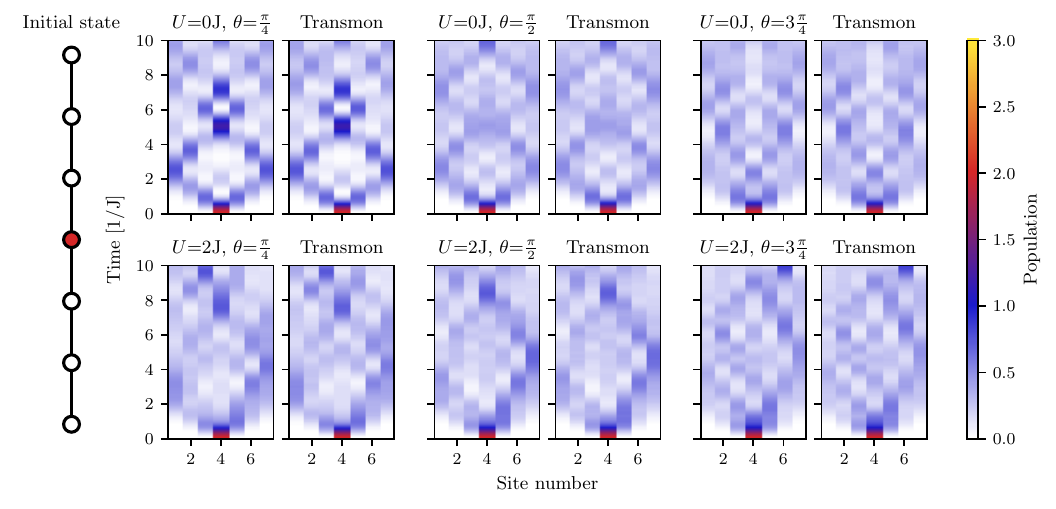}
    \caption{A quantum walk for two interacting anyons initialized on the same site for $U=0$ and $U=2J$ in the top and bottom row respectively. For each interaction strength, the left plot shows the exact dynamics of the AH model, Eq.~(\ref{eq:AHM}), the right plot is the simulated transmon system. We consider the statistical phases $\theta=\pi/4$, $\pi/2$ and $3\pi/4$ for both interaction strengths in the left, middle and right plots respectively. For the non-interacting system, the state delocalizes as the phase increases. For the interacting system, the quantum walk is asymmetric.}
    \label{fig:AHM}
\end{figure*}
To summarize, the recipe for simulating the AH model in Eq.~(\ref{eq:AHM}), where $(\ad_i)^3=0$, is as follows: choose a $\theta$ and a valid set of $p,q,r$, calculate $\gamma$ from $\theta$, fix a base pair $(k_i^{0},k_j^{0})$ in $K^1$, calculate the initial phases $\varphi_0$ and $\varphi_1$, and numerically find the value(s) of $(\lambda_i,\lambda_j)$ that satisfy the constraints. 

\subsection{Excitation manifold and particle filling}
\label{sec:excitations}
The protocols in Secs.~\ref{sec:symmetries} and~\ref{sec:phase} require the truncation $(\ad_i)^4=0$ for the BH model and $(\ad_i)^3=0$ for the AH model. Because the Hamiltonians conserve the particle number, these constraints are automatically satisfied whenever the number of particles in the initial state does not exceed the per-site limit, $\sum_i \langle \n_i\rangle \leq \nmax$.
This condition, while sufficient, is not always necessary. Even when the total particle number exceeds $\nmax$, the protocol remains valid as long as states beyond the truncation threshold do not become populated during the evolution.

In the non-interacting superfluid limit, $U=0$, a system initialized with one particle per site, $\ket{1,1,\dots, 1}$, relaxes locally to a thermal state with a geometric distribution~\cite{cramer2008exact}, for which the probability of finding four or more particles on a site is $P_{\ket{n>3}}\sim6\%$ and finding three or more is $P_{\ket{n>2}}\sim13\%$. However, interactions energetically detune the higher occupied states and suppress their population. In the deep Mott limit, $U\to\infty$, $\ket{1,1,\dots,1}$ is the ground state, and states with two or more particles per site are never populated unless initialized. At intermediate interaction strengths, the validity of $(\ad_i)^4=0$ depends on the strength $U$, the particle density, and the initial proximity of particles. In Appendix~\ref{app:leakage} we show the total population of the forbidden states, $\sum_iP_{{\ket{n>3}}_i}$, for various initial states and interaction strengths. 

Higher excited states are also harder to measure. The experimental fidelity of measuring $\ket{2}$ is lower than that of $\ket{0}$ and $\ket{1}$, and higher states such as $\ket{3}$ may not be measurable at all. Qualitative features like phase transitions may be identifiable even with truncated operators such as $\ket{1}\bra{1}+2\ket{2}\bra{2}$. In our numerical simulations we consider all relevant states to be measurable. 

\section{Numerical simulations}
\label{sec:numerics}

We compare the system of coupled driven transmons to the exact target Hamiltonians. The simulated transmon Hamiltonian is Eq.~(\ref{eq:SC}) where $\omega_i(t)=\omega_i^0+\delta_i\cos(\Omega_i t+\varphi_i)$, with the addition of the counter-rotating terms $g_{ij}(\ad_i\ad_j+\a_i\a_j)$. We have considered the hardware parameters $\omega_i^0/2\pi=4.5$ GHz, $\alpha/2\pi=-250$ MHz, and $g_{ij}/2\pi=6$ MHz for the AH model, and $g_{ij}/2\pi=$ $15$ MHz for the BH model. The simulations for the AH model simulation use a lower capacative coupling strength, because those simulations are more sensitive to RWA errors. See the reported $r_{min}$ and further simulation details in Appendix~\ref{app:simul}. 

Decoherence is not included in our simulations. The evolution time $t = 10/J$ corresponds to approximately $0.5$ $\mu$s for the BH simulations and $1.3$ $\mu$s for the AH simulations, so we expect the system to stay coherent for realistic hardware decoherence times, $T_1,$ $T_2>10$ $\mu$s. The simulations are done using state-vector evolution in \texttt{QuTiP}~\cite{lambert2026qutip}, in a truncated Hilbert space. The truncation level depends on the size of the system and number of particles in the initial state, see Appendix~\ref{app:simul}. In this paper, we only consider the measurement of $\langle\n_i(t)\rangle$. However other observables such as the two-point correlator~\cite{ticea2025observation} or the energy spectrum~\cite{roushan2017spectroscopic} can be measured with transmons, and consequently in our protocol. 

To illustrate our analog simulation protocol, we consider three representative examples: two BH systems with different initial states and one AH system. We vary the interaction strength $U$ in the BH models and both $U$ and the statistical phase $\theta$ in the AH model.

First, we simulate a 2D grid of $4\times3$ sites, with two doublons initialized in two opposite corners, $\ket{\psi_2}=(\ad_1)^2(\ad_{12})^2\ket{0}$ in Fig.~\ref{fig:doublon} for $U=\{0,3J,10J\}$. The system quickly delocalizes when $U=0$, and localizes as $U$ increases. Moreover, the total number of particles exceeds the maximum occupation $\nmax=3$, so the truncation is not guaranteed by particle-number conservation, and for $U=0$ we see in Fig.~\ref{fig:doublon} that the simulation becomes less accurate as time progresses, as discussed in Sec.~\ref{sec:excitations}.  

Second, we consider the BH model for a 1D chain of ten sites, initialized with three excitations, a triplon, in the center site, $\ket{\psi_1}=(\ad_6)^3\ket{0}$, for $U=\{0,2J,4J\}$ in Fig.~\ref{fig:triplon}. We can see that the triplon quickly spreads out across the chain when $U=0$, but stays localized as the interaction is increased. The on-site interaction has a stronger effect on the triplon compared to the doublon, so the system localizes more. The simulated transmon system deviates more from the exact model for larger interaction strengths. The mismatch can be explained by the time-dependent transition amplitudes removed with the RWA, which contribute static occupation-dependent corrections at second order in a high-frequency expansion~\cite{bukov2015universal}.
\begin{figure}[tbp]
    \centering
    \includegraphics[width=\columnwidth]{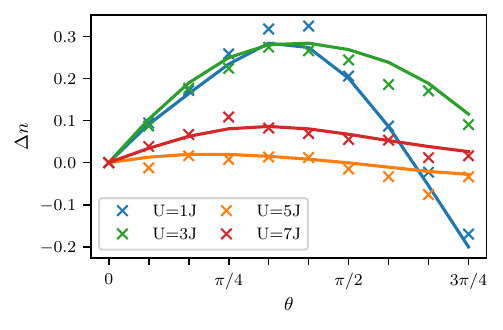}
    \caption{The asymmetry of particles $\Delta n$ taken at $t=5/J$ for various interaction strengths $U$ and statistical phases $\theta$. Solid lines show the exact AH model, Eq.~(\ref{eq:AHM}), and markers show the driven transmons. The transmon system reproduces the position of the asymmetry peak, and its suppression with increasing $U$.}
    \label{fig:deltan}
\end{figure}

Lastly, in Fig.~\ref{fig:AHM} we simulate two anyons initialized in the center of a seven-site 1D chain, $\ket{\psi_3}=(\ad_4)^2\ket{0}$, for $U=0$ (top row) and $U=2J$ (bottom row). For both interaction strengths, we consider three different statistical angles, $\theta=\{\pi/4,\pi/2, 3\pi/4\}$. For $U=0$, the state delocalizes more quickly as the statistical phase $\theta$ increases. At $\theta=\pi/4$, there is some constructive interference at time $t=5/J$, which is not present for the other $\theta$. When the interaction strength is non-zero, the particle walk becomes asymmetric, as predicted in Ref.~\cite{liu2018asymmetric}, and the particles prefer one side, which we can observe clearly in the bottom row of Fig.~\ref{fig:AHM} for all phases $\theta$. To quantify the asymmetry, we consider 
\begin{equation}
    \Delta n = \sum_{i>(L+1)/2}\langle\n_i\rangle-\sum_{i<(L+1)/2}\langle\n_i\rangle,
\end{equation} 
which counts the relative weight of excitation on the right side of the array. We calculate $\Delta n$ at the half-way point $t=5/J$ for $\theta=k\pi/12$, $k\in\{0,\dots,9\}$ and $U=\{1J,3J,5J,7J\}$, shown in Fig.~\ref{fig:deltan}. The transmon system reproduces the $\theta$ dependence of the asymmetry, and its suppression with increasing $U$. 

\section{Conclusions and outlook}
\label{sec:conclusion}

In this paper, we have proposed a flexible quantum analog simulation protocol that realizes interacting anyons and bosons with model parameters that can be varied over a broad continuous range, such as the interaction strength $U$ and the anyonic statistical phase $\theta$, using only capacitively coupled flux-tunable transmons. To our knowledge, this constitutes the first proposal for analog simulation of the anyon–Hubbard model on a superconducting platform. For the Bose–Hubbard model, our protocol supports up to three particles per site and does not require a separate constraint on the total particle number, apart from the limit set by the maximum site occupancy. In contrast, a previous transmon realization is restricted to two particles in total, regardless of the number of sites~\cite{wang2026observing}.
Our protocol can furthermore target correlated, density-dependent hopping operators, while previous experimental realizations are limited to standard, density-independent hopping.

These capabilities all stem from a single mechanism. Driving the frequencies of neighboring transmons sinusoidally, in an alternating manner, makes each many-body transition simultaneously resonant with multiple frequency sideband combinations. The effective transition amplitude is the sum of the sideband amplitudes. In a rotated frame, the on-site interaction becomes $U=\alpha-h$ and is tunable with the frame parameter $h$. The resonance condition in this frame depends on $h$, which in turn fixes the drive frequencies $\Omega_i$. By limiting the excitation manifold to three particles per site, $(\ad_i)^4=0$, we have enough experimental parameters to tune the ratios between the relevant transition amplitudes, and match them to the corresponding hopping amplitudes of a target model. We implement the standard BH model by setting all transition amplitudes equal, and can implement more general correlated hoppings by choosing the transition amplitudes independently. Furthermore, by driving the transmons with phases, each sideband amplitude is complex-valued, giving transition-dependent phases. For two particles per site $(\ad_i)^3=0$, we can engineer the drive phases to create density-dependent Peierls phases, simulating anyonic dynamics described by the 1D AH model with tunable interaction $U$ and statistical phase $\theta$. 

We have validated our protocol numerically by simulating the driven transmon array with experimentally realistic parameters, and comparing the dynamics to those of the exact target models. The simulated system reproduces the interaction-dependent localization of a triplon in a 1D chain and two doublons on a 2D grid for the BH model, as well as the asymmetric quantum walk of two interacting anyons in the AH model. Here, the transmon system captures the assymetric quantum walk, and its dependence on the statistical phase $theta$, including its peak position and suppression with increasing interaction strength.

The parameters used in our simulations are typical of current superconducting devices, and the protocol requires only frequency modulation of standard flux-tunable transmons, so we expect it to be implementable on existing hardware. A key experimental requirement is accurate calibration of the individual drive amplitudes. In future work, we would like to expand the protocol by considering higher-order harmonics of the flux frequency, either to target more advanced Hamiltonians or to tune additional transition amplitudes, further increasing the number of particles per site beyond what we reach here. In the same spirit, when the target model does not have complex transition amplitudes, it may be possible to leverage the drive phases to tune a larger number of transitions. 

Finally, the simultaneous and independent control of the on-site energy, the interaction, and the density-dependent hopping amplitudes suggests that flux-driven transmon arrays can realize a considerably broader class of lattice models than those treated here, and we encourage the search for further realizable target models.

\begin{acknowledgments}
We acknowledge useful discussions with Erik van Loon, Théo Sépulcre, Simon Petterson Fors,  Alberto Del Ángel, and Anuj Aggarwhal. We acknowledge support from the Knut and Alice Wallenberg Foundation through the Wallenberg Centre for Quantum Technology (WACQT). L.G.-Á. further acknowledges funding under Horizon Europe programme HORIZON-CL4-2022-QUANTUM-01-SGA via the project 101113946 OpenSuperQPlus100.
\end{acknowledgments}

\appendix

\section{Phases removable with a diagonal frame}
\label{app:phase}

When all sums of Bessel pairs $k_i+k_j$ in $K^m$ are odd, the transition amplitudes are related as $G^m=-G^{-m}$, as shown in Sec.~\ref{sec:symmetries}, which we interpret as a $\pi$ phase between those transitions. Depending on the value of $m$, the number of particles in the system, $N_\text{total}$, and the per-site particle limit, $\nmax$, this $\pi$ phase may be absorbed into a diagonal frame that leaves the population dynamics $\langle n_i(t)\rangle$ unchanged. In this appendix, we determine which configurations of $m$, $N_\text{total}$ and $\nmax$ give removable phases. The results are summarized in Table~\ref{tab:phases}.

Since the drive parameters alternate across the lattice, the sets of Bessel order pairs also alternate, with flipped signs. That is, for coupled transmons $i$ and $j$, if $(k_i,k_j) \in K^m_{ij}$, where the subscript indicates the transmon pair, then for the coupled transmons $j$ and $\ell$, with $\ell \neq i$, the pair $(-k_j,-k_i)$ lies in $K^{-m}_{j\ell}$. Let $|\psi_1\rangle$ be the final state of the $m$ transition between $i$ and $j$, for example $|\psi_1\rangle\langle\psi_0|$. The $-m$ transition between $j$ and $\ell$ then starts from that same state, for example $|\psi_2\rangle\langle\psi_1|$. Because both transitions involve $|\psi_1\rangle$, the $\pi$ phase between them can be absorbed into a phase on that state, $|\psi_1\rangle \to -|\psi_1\rangle$, which leaves the observable $\langle n_i(t)\rangle$ unchanged.

Let us consider an example with three transmons and two particles. A $\pi$ phase between the $\pm 1$ transitions between the first and second transmon gives the operator $\ket{200}\bra{110}-\ket{110}\bra{020}$. The $\pi$ phase means that the sums $k_i+k_j$ of the pairs in $K^{1}_{1,2}$ are odd, and hence that the sums of the pairs in $K^{-1}_{2,3}$ are odd as well. The $\pm 1$ transitions between the second and third transmon therefore carry the opposite relative sign, $-|020\rangle\langle 011| + |011\rangle\langle 002|$. Taking the frame that transforms $|2\rangle_2 \to -|2\rangle_2$, we obtain $|200\rangle\langle 110| + |110\rangle\langle 020| + |020\rangle\langle 011|+ |011\rangle\langle 002|$, i.e., no relative phases between the operators.

However, the frame also adds a $\pi$ phase to all other transition containing $|\psi_1\rangle$ ($\ket{2}_2$ in our previous example). If $|\psi_1\rangle$ holds all $N_\mathrm{total}$ particles of the system, there is only one such transition, $|N,0\rangle\langle N-1,1| + \mathrm{h.c.}$, so a phase on $m = N_\mathrm{total}-1$ can always be removed. Otherwise we must check how the frame affects the remaining transitions containing $|\psi_1\rangle$. To do this we consider a general sign-flipping frame on two sites, $\sum_{n_i,n_j} A_{n_i}|n_i\rangle\langle n_i| \otimes B_{n_j}|n_j\rangle\langle n_j|$, with $A_{n_i}, B_{n_j} = \pm 1$.

We now work through one explicit configuration, a system with three or more particles where each site holds at most two particles. The remaining configurations follow from the same argument and are collected in Table~\ref{tab:phases}. For $N_\mathrm{total} \geq 3$ and $\nmax = 2$, the transitions transform as
\begin{align*}
|10\rangle\langle 01| &\to A_1A_0B_0B_1 |10\rangle\langle 01|
  = F_1 |10\rangle\langle 01|, \\
|20\rangle\langle 11| &\to A_2A_1B_0B_1 |20\rangle\langle 11|
  = F_2 |20\rangle\langle 11|, \\
|11\rangle\langle 02| &\to A_1A_0B_1B_2 |11\rangle\langle 02|
  = F_3 |11\rangle\langle 02|, \\
|21\rangle\langle 12| &\to A_2A_1B_1B_2 |21\rangle\langle 12|
  = F_4 |21\rangle\langle 12|.
\end{align*}
Since $A_{n_i}^2 = B_{n_j}^2 = 1$, we have $F_4 = F_1F_2F_3$. Both $F_1$ and $F_4$ describe the transformation of $m=0$ transitions, so $F_1 = F_4$, and therefore $F_2F_3 = 1$. As $F_2$ and $F_3$ are both $\pm 1$, this gives $F_2 = F_3$: the $m=1$ and $m=-1$ transitions must transform in the same way, and any $\pi$ phase between them cannot be removed. Following the same reasoning for the remaining cases with $N_\mathrm{max} \leq 3$ gives the entries of Table~\ref{tab:phases}. 
\begin{table}
    \caption{\label{tab:phases} Classification of the removable $\pi$ phases between the $\pm m$ transitions are removable for different $N_\text{total}$ particles in the system with at most $\nmax$ particles per site.}
    \begin{tabular}{C{0.23\columnwidth }C{0.23\columnwidth }C{0.23\columnwidth }C{0.23\columnwidth }}
    \hline\hline
         Removable&$N_{\rm{total}}$  &$\nmax$ & $m$ \\
         \hline
        Yes&2 &2  &0 \\
        Yes & 2 & 2 &$\pm 1$\\
        Yes &$\geq3$ & 2 &0 \\
        No & $\geq3$ & 2 & $\pm 1$\\
        Yes & 3& 3& 0\\
        No & 3& 3& $\pm 1$\\
        Yes & 3& 3&$\pm 2$\\
        No & $\geq4$& 3&0\\
        No & $\geq4$ & 3 & $\pm 1$\\
        No & $\geq4$& 3& $\pm 2$\\
        \hline\hline
    \end{tabular} 
\end{table}

\section{Numerical simulation details}\label{app:simul}
Here we give a summary of all simulation parameters and numerical truncations used to generate the figures in the main text.

For the reported values of worst-case RWA ratio from Eq.~(\ref{eq:rmin}), $r_\text{min}$ we set the frame parameter $h = \alpha$. We consider multiple drive frequencies $(\Omega_i,\Omega_j)=(ph/r,qh/r)$, where $p+q$ is an even number and $p,q\leq 7$, and numerically search for the amplitude and frequency ratios $(\lambda_i,\lambda_j)$ that implement the desired target model, and choose the configuration with the highest $r_{\text{min}}$. We truncate the Bessel orders in the transition amplitudes to $|k_i|\leq 10$.

When simulating the BH model we used $\Omega_i=3h/2$ and $\Omega_j=5h/2$, with $\lambda_i =  0.947$ and $\lambda_j= 2.890$, giving $G^m/g_{ij}=0.172$, and $r_\text{min}=233$. We set the static parameters to $\omega_i^0/2\pi=4.5$ GHz, $\alpha/2\pi=-250$ MHz, and $g_{ij}/2\pi=15$ MHz. The AH model requires one setup per phase $\theta$, each described in Table~\ref{tab:ahmparam}. The drive frequencies $\Omega_i$ used for the AH model give rise to the $\pi$ phase discussed in Sec.~\ref{sec:symmetries}, but as shown in Appendix~\ref{app:phase}, for a system of two particles the $\pi$ phase does not change the dynamics. For the simulation of the AH model we used $g_{ij}/2\pi=6$ MHz, as we found it to be more sensitive to RWA errors, see the reported $r_\text{min}$ in Table~\ref{tab:ahmparam}. The hilber space of the transmon simulation is truncated to exclude all states with more than $N_{\text{total}+3}$ excitations. The exact models conserve particle number, so those simulations only require a subspace of states with the same particle number as the initial state. 

\begin{table}[htbp]
    \caption{\label{tab:ahmparam} Summary of the drive frequencies $(\Omega_i,\Omega_j)=(p/r,q/r)$, drive amplitudes $\lambda_i=\delta_i/\Omega_i$, phase parameter $\gamma$ and boundary drive phases $\varphi_0,\varphi_1$ used to simulate different statistical phases $\theta$ of the AH model in Eq.~(\ref{eq:AHM}).}
    \begin{tabular}{C{0.1\columnwidth }C{0.15\columnwidth }m{0.22\columnwidth }m{0.34\columnwidth }C{0.1\columnwidth }}
    \hline\hline
         $\theta$&$(p,q,r)$  &$(\lambda_i,\lambda_j)$ & $(\gamma,\varphi_0,\varphi_1)$ & $r_\text{min}$\\
         \hline
        $\pi/12$& (1, 3, 1)& (1.29, 3.34)& (0.785, -1.44, -5.11) & 228\\
        $2\pi/12$& (1, 3, 1)& (1.41, 3.29)& (-1.57, -0.262, 0.785) & 224\\
        $3\pi/12$& (1, 3, 1)& (1.78, 3.21)& (2.36, 1.96, 3.53) & 222\\
        $4\pi/12$& (7, 1, 1)& (0.88, 6.29)& (-95.3, -91.6, 0.524)&177\\
        $5\pi/12$& (1, 3, 1)& (3.03, 1.50)& (-3.93, -8.51, -21.6)&153\\
       $6\pi/12$& (1, 3, 1)& (3.26, 0.69)& (-4.71, -10.2, -25.9)&98\\
       $7\pi/12$& (3, 1, 1)& (1.36, 6.21)& (-5.50, -12.2, -2.23)&210\\
        $8\pi/12$& (1, 7, 1)& (6.55, 0.836)& (-102, 2.09, 117)&161\\
        $9\pi/12$& (1, 3, 1)& (5.41, 1.14)& (-30.6, -1.96, 24.7)&174\\
        \hline\hline
    \end{tabular}
\end{table}

\section{Leakage to higher excited states}
\label{app:leakage}
\begin{figure*}
    \centering
    \includegraphics[width=\textwidth]{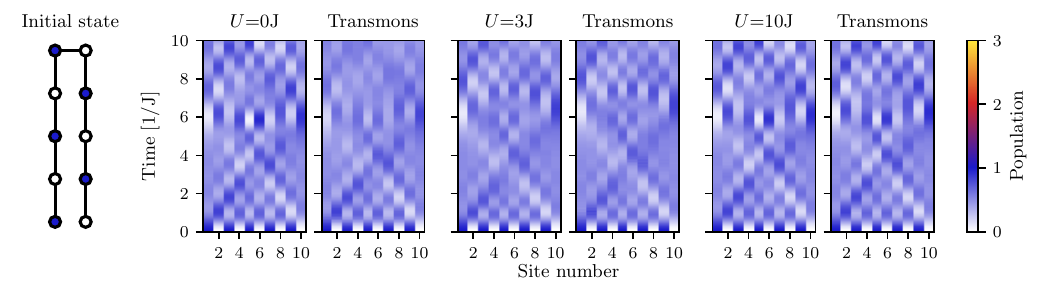}
    \caption{A quantum walk of a half-filled initial state on a ten-site 1D chain for $U/J=\{0,3,10\}$. For each interaction strength, the left plot shows the dynamics of exact BH model, Eq.~(\ref{eq:BH}), and the right plot, the simulated transmon system. The system contains a total of five particles, which exceeds the maximum occupation per site $\nmax=3$, so the truncation is not guaranteed by particle-number conservation.}
    \label{fig:halffilling1d}
\end{figure*}
\begin{figure}
    \centering
    \includegraphics[width=\columnwidth]{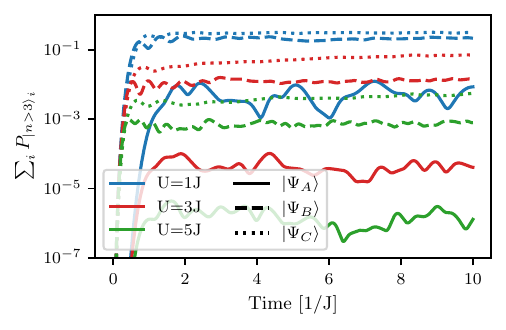}
    \caption{Total leakage probability to the forbidden states $\sum_iP_{\ket{n>3}_i}$ for a 1D array of eight sites, and three different initial states: $\ket{\Psi_A}$, $\ket{\Psi_B}$ and $\ket{\Psi_C}$, described in Appendix~\ref{app:leakage}. The number of particles is four for $\ket{\Psi_A}$, and eight for $\ket{\Psi_B}$ and $\ket{\Psi_C}$.}
    \label{fig:leakage1d}
\end{figure}
\begin{figure}
    \centering
    \includegraphics[width=\columnwidth]{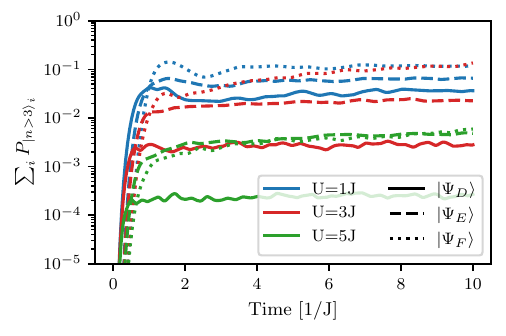}
    \caption{Total leakage probability to the forbidden states $\sum_iP_{\ket{n>3}_i}$ for a 2D array of $4\times3$ sites, and three different initial states: $\ket{\Psi_D}$, $\ket{\Psi_E}$, and $\ket{\Psi_F}$, described in Appendix~\ref{app:leakage}. All three states have six particles.}
    \label{fig:leakage2d}
\end{figure}

Here, we study how strongly the system leaks into the forbidden states $\ket{n>3}_i$, corresponding to local occupation numbers of four or more, and its dependence on the interaction strength and initial state. We first compare the driven transmon array to the exact BH model for a quantum walk of five particles, in a ten-site 1D chain where every other site is initialized with a particle (see Fig.~\ref{fig:halffilling1d}). For $U = 0$, there is no energetic suppression of highly occupied states, so the leakage is expected to be larger. Consistently, we observe a growing drift between the BH model and transmon system that increases with time. For $U = 3J$ and $U=10J$, the agreement between the two systems improves significantly. 

To quantify this leakage, we consider the probability of occupying states with more than three particles on a site $i$, $\sum_iP_{\ket{n>3}_i}$, as discussed in Sec.~\ref{sec:excitations}. We evaluate this quantity for the exact BH model across a range of interaction strengths and initial states. In Fig.~\ref{fig:leakage1d}, we show the results for an eight-site 1D array. We use three initial states: one with half the sites initialized with one particle $\ket{\Psi_A}=\ad_1\ad_3\ad_5\ad_7\ket{0}$, one with all sites initialized with one particle $\ket{\Psi_B}=\prod_i\ad_i\ket{0}$, and one with half the sites initialized with two particles $\ket{\Psi_C}=(\ad_1)^2(\ad_3)^2(\ad_5)^2(\ad_7)^2\ket{0}$. We find that the leakage is significantly higher already at $U=3J$ for the unit-filled states, i.e., those with one particle per site on average, $\ket{\Psi_B}$ and $\ket{\Psi_C}$. In contrast, for the half-filled state, $\ket{\Psi_A}$, the leakage remains lower for each interaction strength $U$. 

Additionally, in Fig.~\ref{fig:leakage2d} we show the leakage probability $\sum_iP_{\ket{n>3}_i}$ for a 2D system of $4\times 3$ sites. We consider three initial states with the same total number of particles: one where half the sites are initialized with one particle $\ket{\Psi_D}=\ad_1\ad_3\ad_6\ad_8\ad_9\ad_{11}\ket{0}$, one where a quarter of the sites are initialized with two particles $\ket{\Psi_E}=(\ad_1)^2(\ad_8)^2(\ad_{10})^2\ket{0}$, and one where a sixth of the sites are initialized with three particles $\ket{\Psi_F}=(\ad_1)^3(\ad_{12})^3\ket{0}$. As in the 1D case, we find that increasing the interaction strength $U$ suppresses leakage. Moreover, initial states with a stronger local concentration of particles exhibit larger leakage probabilities, even when the total particle number is fixed.

\clearpage
\bibliography{references}

\end{document}